# On the calculation of equilibrium thermodynamic properties and the establishment of statistical-thermodynamically-consistent finite bound-state partition functions in nonideal multi-component plasma mixtures within the chemical model

Mofreh R. Zaghloul

Department of Physics, College of Sciences, United Arab Emirates University,

P.O.Box 17551, Al-Ain, UAE.

**ABSTRACT**

The problem of the calculation of equilibrium thermodynamic properties and the establishment of statistical-thermodynamically-consistent finite bound-state partition functions in nonideal multi-component plasma systems is revised within the chemical picture. The present exploration accompanied by the introduction of a generalized accurate formulation, in terms of the solution of the inverse problem, clears ambiguities and gives a better understanding of the problem on top of pointing out weaknesses and inaccuracies/inconsistencies buried in widely used models in the literature.

## 1 – Introduction

Within the chemical picture and for a fixed volume and constant temperature the Helmholtz free energy function is commonly used as the thermodynamic potential that can fully describe the thermodynamic properties of the assembly. Under equilibrium conditions this thermodynamic potential is minimized. In the early 1960s Gilda Harris et al. [1] introduced the approach of free energy minimization for the calculation of the equation-of-state and thermodynamic properties for different complex plasma systems. Since then the free energy minimization procedure has become widely used for the calculation of the equation-of-state, thermodynamic properties and radiative characteristics of laboratory and space plasmas [1-5]. At high densities, interactions and coupling mechanisms among different species in the system bring about a departure from the ideal behavior. This departure from the ideal behavior is usually taken into account in terms of a corrected or modified free energy function. According to Hummer & Mihalas (HM) [2] and Potekhin [4], an appealing feature of the free-energy minimization method is the assumed factorizability of the translational, configurational and internal components of the total many-body partition function and the corresponding separability of the components of the Helmholtz free energy such that

$$F(V, T, \{N_{j,\zeta}\}) = F_{trans} + F_{config} + F_{int} \qquad (1)$$

where $V$ is the volume, $T$ is the absolute temperature and $N_{j,\zeta}$ refers to the population of the species of ionization state $\zeta$ of the chemical element $j$ in the system. The terms $F_{trans}$, $F_{config}$ and $F_{int}$ refer, in order, to the total translational, configurational and internal components of the free energy. As indicated by Potekhin [4] the internal structure of a composite particle is generally affected by the surrounding and therefore the separation in (1) is approximate.

One useful feature of the chemical picture is that it facilitates introduction of the quantum-mechanical results, obtained for isolated bound species, into the thermodynamic plasma model. However, some problems arise because collective quantum properties of the plasma are not identical to the sum of the corresponding properties of the isolated species. In particular, strong particle interactions affect the bound state structure what manifests itself in the dependence of the internal partition functions on density [6] and one has to deal with these plasma nonideality effects (interparticle interactions) on the internal partition functions of bound species. Therefore the chemical picture entails inevitable approximations, which are usually treated by more or less heuristic methods. Even though, the chemical model (or free energy minimization) is the most conventional choice, since its canonical conjugate variables (volume, temperature and particle number) are quantities that can be directly controlled in laboratory experiments.

Although the physical model or "physical picture" of building the plasma up from clusters of elementary particles provides a consistent alternative to the calculation of the thermodynamic functions (see for example [7-9]), a lot of difficulties appear connected with its implementation for multi-component plasma systems and, in particular, for plasma which contains atoms and ions of several chemical elements [6]. It was concluded by W. Ebeling [10] that *the most realistic description is given by advanced chemical descriptions although still connected with open problems*.

The minimization of the free energy function subjected to the constraints of electroneutrality and conservation of nuclei is known to assure thermodynamic consistency among the population numbers and different thermodynamic functions derived from the same free energy function. Two approaches for the minimization of the free energy function have been used in the literature; 1- using numerical optimization algorithms [1-3] and 2- by casting the minimization equations analytically into the form of nonideal (modified) Saha equations and solving the resulting set of nonlinear algebraic equations subjected to the above mentioned constraints [4,11,12]. It has to be noted, however, that a *consistent* scheme for establishing finite bound-state partition functions in the plasma environment is required for both methods of solution. Such a scheme should formally mimic the mechanism by which bound states are swept from being "bound" to the "continuum" causing the well-known phenomenon of *pressure ionization* at high densities.

The use of the temperature-dependent Planck-Larkin partition function (PLPF), for nonideal plasma systems has been critically criticized by Rouse in Ref. [13] who argued that the use of PLPF for nonideal plasma systems suffers major physical problems. In a response to Rouse's criticism, Ebeling et al in Ref. [14] recommended the use of the discrete energy states of the Bethe-Salpeter equation (BSE) in the PLPF, for quantum statistical Coulomb system with bound states, in order for the latter to become temperature-and density-dependent. Following to Ebeling's et al response to Rouse's criticism, Rogers in Ref. [7] and Däppen et al (see, for example Refs [15, 16]) stressed that despite its name the PLPF is not a true internal partition function but merely an auxiliary term in a virial coefficient. Accordingly, the present analysis and discussion will be concerned with and restricted to formulations of the internal partition functions which are both temperature and density-dependent within the chemical picture. The manifestation and/or interpretation of the phenomenon of pressure ionization is therefore related to the scheme used for establishing finite bound-state internal partition functions.

## 2– Internal Partition Function and the Occupational Probability Formalism

As indicated above, the central problem of thermodynamic equilibrium at high density, within the chemical picture, lies therefore in the consistent determination of the density and temperature dependence of the scheme used to formulate finite internal partition functions of the ensemble of bound states. The cutoff of the bound-state partition function at some maximum state-dependent energy level in order to account for the effects of the environment (see for example Refs [2,4,17-21]) is known to abruptly switch a state from being "bound" to "free". This abrupt disappearance of the bound state into the continuum will lead to discontinuities and singularities in the free energy and its derivatives in the expressions of the thermodynamic functions. In order to avoid these discontinuities in the internal partition functions, Hummer & Mihalas (HM) [2], following a historical work by Fermi [22], suggested the assignment of *weights* or *occupational probabilities* to all bound states of all species. These occupational probabilities depend, in general, on the occupation numbers of the species and the internal partition function (IPF) can therefore be written as

$$Q_{\text{int},j,\zeta}(\rho,T) = Q_{\text{int},j,\zeta}(V,T,\{N_{j,\zeta,i}\}) = \sum_i w_{j,\zeta,i}(V,T,\{N_{j,\zeta,i}\}) \exp\left(\frac{-\varepsilon_{j,\zeta,i}}{K_B T}\right)$$

$$= \sum_i g_{j,\zeta,i}\, \omega_{j,\zeta,i}(V,T,\{N_{j,\zeta,i}\}) \exp\left(\frac{-\varepsilon_{j,\zeta,i}}{K_B T}\right)$$

(2)

where $\rho$ is the mass density, $K_B$ is the Boltzmann constant, $N_{j,\zeta,i}$, $g_{j,\zeta,i}$ and $\varepsilon_{j,\zeta,i}$ represent, in order, the occupation number, the statistical weight and the excitation energy above the ground state for the excited state *i* of the ion $\zeta$ of the chemical element *j* in the mixture. The factor $w_{j,\zeta,i}(V,T, \{N_{j,\zeta,i}\}) = g_{j,\zeta,i}\, \omega_{j,\zeta,i}(V,T, \{N_{j,\zeta,i}\})$ is the corresponding state-dependent occupational probability of that level. The occupational probability, $w_{j,\varsigma,i}(V,T, \{N_{j,\varsigma,i}\})$, is presumed to decrease continuously and monotonically as the strength of the relevant interactions increases in order to produce a physically reasonable continuous transition between bound and free states. Further, the occupational probability $w_{j,\varsigma,i}(V,T, \{N_{j,\varsigma,i}\})$ or equivalently the weight $\omega_{j,\varsigma,i}(V,T, \{N_{j,\varsigma,i}\})$ should drop strongly to zero as the binding energy of a level below the unperturbed continuum goes to zero in order to provide natural and smooth truncation of the internal partition function. The approach of the occupation probability formalism is extensively adopted and used in the literature (see for example Refs [23-27]). However, as shown by Fermi [22] and emphasized by HM [2] and Potekhin [4], the introduction of an occupational probability $w_{j,\varsigma,i}(V,T, \{N_{j,\varsigma,i}\})$ is equivalent to a modification of the free energy such that $w_{j,\varsigma,i}(V,T, \{N_{j,\varsigma,i}\})$ becomes consistent with and can be derived from the adopted form of the $F_{config}$.

Both of HM and Potekhin assumed the separability of the configurational component of the free energy and possibility of deriving corresponding occupational probabilities, $w_{j,\varsigma,i}(V,T, \{N_{j,\varsigma,i}\})$, that depend on the occupation numbers of the individual excited states; an assumption that will be logically and analytically disproved in the present paper.

It has to be noted that HM [2] considered the simplest case of a single species *perfect* gas of neutral particles where the ionization processes are neglected and consequently no charged particles are involved. They assumed a separable configurational component, $f$, of the free energy that is supposed to depend explicitly on the occupation numbers of the individual excited states $\{N_i\}$ and allegedly assumed to be responsible for the truncation of the internal partition function. Considering the equilibrium between excitation/de-excitation processes they derived an expression for the occupational probability which takes the form

$$w_i = g_i \, \omega_i \equiv exp\left(-\frac{\partial f / \partial N_i}{K_B T}\right) \qquad (3)$$

Even though, using this form of the occupational probability and substituting back into the free energy function they recovered a residual term

$$\left[f - \sum_i N_i (\partial f / \partial N_i)\right] \qquad (4)$$

which indicates that such a simple, separability of the configurational component of the free energy, in the case of using an occupational probability in the form (3), is not always possible. As commented by HM, the existence of the residual term expressed by Eq. (4) has been known for many years; however, many authors using heuristic occupation probabilities have ignored it and therefore arrived at formulations that are statistical-mechanically inconsistent. In order to assure thermodynamic consistency and to be able to use the relation for $w_i$ as in Eq. (3) Hummer & Mihalas had to assume linear dependence or astute linearization of the interaction term on $\{N_i\}$ and they made every possible approximation in order to satisfy such a linear dependence even with the inclusion of ionization and the involvement of the long range Coulomb forces among the charged particles. Moreover, the use of $w_i$ as given by HM and *unperturbed energy levels* in the calculation of HM partition functions does not exclusively ensure convergence and finiteness of the IPF nor the manifestation of the phenomenon of pressure ionization; a problem that appears clearly at high temperatures.

On the other side, Potekhin [4] considered the case of hydrogen plasma including charged particles (i.e; taking ionization processes into consideration) and assuming the separability of the configurational free energy, the author derived a form for the occupational probability as

$$w_i \equiv g_i \, exp\left( -\frac{1}{K_B T} \left( \frac{\partial F_{config}}{\partial N_i} - \frac{\partial F_{config}}{\partial N_p} - \frac{\partial F_{config}}{\partial N_e} \right) \right) \qquad (5)$$

where $N_p$, $N_e$, and $N_i$ are the occupation numbers of protons, electrons and hydrogen atoms at excited state *i*, respectively. Eq. (5), which reduces to Eq. (4) in the absence of charged particles, was introduced as a form of the occupational probability that is thought to be consistent with and can be derived from the proposed form of a separable configurational component of the free energy function. It has to be noted that Hummer & Mihalas in Ref. [2] split the configurational free energy into two components; a component *f* which is supposed to depend explicitly on the populations of the individual excited states and hence is supposed to be responsible for the truncation of the internal partition function and the component, **F₄**, accounting for the Debye-shielded Coulomb interactions among free charges (as defined by GHR in Ref. [28]). In their description of the occupation probabilities, HM in Ref. [2] stated that "In addition, they can be incorporated in a *statistical mechanically* consistent way (see **§** II) into the free energy of the plasma **along with** the best-available models for other nonideal effects ***not explicitly accounted for in the occupation probabilities***". Such splitting is not made in Ref. [4] where both components are combined into the term *F$_{config}$* and hence this term is supposed to be responsible for the truncation of the partition function in Ref. [4].

In the present study we show that the use of a separable configurational component of the free energy function, equivalent to a form of *w$_i$* similar to those derived by HM or Potekhin, is *strictly conditional* for a very limited situations of little or no practical importance as far as thermodynamically-consistent establishment of bound-state partition functions is concerned. Further, we show that assuming a separablility of the configurational component of the free energy function cannot lead to a truncation of the IPF unless an abrupt cutoff scheme is adopted. In addition we introduce, as a remedy, a generalized more accurate formulation of the problem in terms of the solution of the inverse problem in which, based on physical bases, an occupational probability (or more precisely, a scheme to establish finite IPFs) is prescribed in advance to satisfy and to assure a smooth *truncation* of the IPF and a manifestation of the pressure ionization. The resulting IPF in this case can then be used to calculate the occupation numbers and to derive the corresponding set of modified thermodynamic functions without the need to explicitly construct a separable form for the term of the configurational component, allegedly responsible for the truncation of the IPF, in the Helmholtz free energy function. This proposed solution stands on a fundamental statistical thermodynamic principle that is "all major thermodynamic functions could be expressed in terms of the partition function and once the partition function has been evaluated in terms of molecular quantities including molecular interactions, the equilibrium behavior of matter will be completely understood" [29-31].

Values of the corrected or modified free energy of the system, if needed, can therefore be found either analytically or numerically depending on the used form of the occupational probability (or the scheme used to establish finite IPF).

## 3– Weaknesses and Inconsistencies of HM Occupational Probability Formalism

In the present section we point out the weaknesses and inconsistencies of the HM occupational probability formulation used in the calculation of equation-of-state of nonideal multi-component plasma systems in the Opacity-Project (OP). The inconsistencies in Potekhin's formulation are pointed out in another place [32].

Before going deeper into the analysis it has to be noted that the inconsistencies in Refs. [2,4] are very clear and simply recognizable from the assumption of the factorizabiliy of the partition function or equivalently from the assumption of separability of the free energy components. Simply, the factorizability of the partition function (or equivalently the separability of the free energy components) implies that various types of energies are independent of each other (the uncoupling assumption adopted in Refs. [2,4]). The separation of the configurational free energy therefore indicates that it has no influence on the internal free energy component. Now, if the configurational free energy (or interaction energy) has no influence on the internal free energy, the expectation that including a separable configurational component could lead to a truncation of the internal partition function is physically and logically incorrect because they are independent of each other by assumption. Unfortunately this simple logical fact was not recognized by the authors of Refs. [2,4] or others who used their formulations. It has also to be noted that the problems with HM model [2] are manifold and they have observational consequences, however, our interest here is to show the inconsistencies in its mathematical frame work. Other inadequacies of HM model including errors resulting from a simplified microfield treatment (a power law fit to a Holtsmark microfield distribution), the lack of higher-order Coulomb terms (beyond the clearly outdated $\tau$-correction of the Debye-Hückel approximation, and the absence of a non-linear configuration term for neutral extended species have been shown in other places [33,34,35].

Considering a single-species (monatomic) *perfect* gas (no ionization or dissociation), Hummer & Mihalas [2], expressed the entropy of the gas as

$$S = K_B N \left( \frac{3}{2} \ln T + \ln V + \ln G + \frac{5}{2} \right) - K_B \sum_i N_i \ln N_i \qquad (6)$$

where $G = (2\pi m K_B / h^2)^{3/2}$. In the same time they used the following expression for the internal energy

$$U = \frac{3}{2} N K_B T + \sum_i N_i \varepsilon_i \qquad (7)$$

where $\{N_i\}$ are the occupation numbers satisfying the constraint of constant number of molecules;

$$N = \sum_i N_i = constant \qquad (8)$$

Ahead of proceeding to point out some of the weaknesses and limitations of the HM occupational probability formalism, it may be enlightening to discuss some fundamental aspects regarding Eqs. (6) and (7) used by HM to represent the entropy and internal energy of a monatomic perfect gas:

Firstly, Eq. (6) may correctly and appropriately represent the entropy of a perfect monatomic gas if the subscript *i* is dropped from the last term in the right-hand-side (entropy of mixing). In such a case the last term becomes $-K_B N \ln N$ and Eq. (6) becomes in complete agreement with Eq. (43.5) in Landau and Lifishtiz [29] for atoms presumed to be in their *non-degenerate* ground states (i.e., no excitation and no degeneracy of the ground state as well)[1]. Secondly, the inclusion of the *excitation energies* (excitation/de-excitation interactions) in Eq. (7) breaks the original assumption of a monatomic perfect gas, by definition, where a monatomic perfect gas has only translational kinetic energy (no means of storing energy except as kinetic energy). Therefore Eq. (6) and Eq. (7) are not compatible. This incompatibility is undoubtedly reflected in the free energy expression presented in their work as well; that is

$$F = -K_B T N \left(\frac{3}{2} \ln T + \ln V + \ln G + 1\right) + \sum_i N_i \varepsilon_i + K_B T \sum_i N_i \qquad (9)$$

It may be recalled here that as indicated by HM "It should be emphasized, however, that in using the free energy method the statistical mechanical consistency of the formulation has to be checked because the method will obligingly produce results for *any* well-behaved *F*, even if that model of the free energy is incompatible with statistical mechanics (or is even physically nonsensical!). In order to explain this fundamental point more, let us consider, for the moment, a first interpretation in which one sticks to the original assumption of a monatomic perfect gas that HM started with bearing in mind that the last term in the entropy expression given in Eq. (6) should be $-K_B N \ln N$. In that case no excitation energies should be included and the terms including these excitation energies should be dropped out from Eq (7) and Eq (9) in order for these equations to correctly and compatibly represent a monatomic perfect gas. On the other hand, if one chooses to interpret the HM treatment as if it were pertaining to a monatomic gas with different excitation energies assuming the applicability of the Boltzmann distribution for excitation energies at low densities, then Eq. (7) is correct, however, Eq. (6) must include the entropy due to the degeneracy of the energy levels and an associated term has to be included in Eq. (9) as well. This latter interpretation may be supported by the appearance of the excitation energies in Eqs. (6,7 and 9).

In order to explicitly identify and put a figure on these missing terms in the HM treatment, for such an interpretation, we start using the general expression for the entropy of any material [29-31]

$$S = K_B N \left[ \ln\left(\frac{Q_{tot}}{N}\right) + \frac{U}{NK_B T} + 1 \right] \qquad (10)$$

where $Q_{tot}$ is the total partition function of the molecule. Equation (10) is valid for any state of matter, solid, liquid or gaseous [31]. Assuming factorizability of the translational and internal components of the many-body partition function as adopted by HM in Ref. [2] and Potekhin in Ref. [4] the total partition function of the molecule is given by the product of the translational partition function, $Q_{trans}$, and the internal partition function, $Q_{int}$; that is $Q_{tot} = Q_{trans} Q_{int}$. Needless to say that any factorizable configurational partition function, if any, can be represented by a factor to be multiplied by and hence can be combined with the IPF. However, the opposite may not be always possible that is the total partition function can not always be factorized unless the assumption of the uncoupling of the translational, internal and interaction energy components is adopted. Substituting from Eq. (7) into Eq. (10), for such a case, expanding the logarithm and making use of the constraint of constant number of molecules (Eq. (8)) one gets

---

[1] The chemical constant $\zeta$ in Landau's and Lifishtiz equation corresponds to the factor *G* in Eq. (6); see also Eq. (45.4) in Ref. [29].

$$S = K_B N \left[ \ln Q_{trans} + \ln Q_{int} - \ln N + \frac{1}{NK_B T} \sum_i N_i \left( \frac{3}{2} K_B T + \varepsilon_i \right) + 1 \right]$$

$$= K_B \sum_i N_i \times \left[ \ln Q_{trans} + \ln \frac{Q_{int}}{N} + \frac{5}{2} + \frac{1}{NK_B T} \sum_i N_i \varepsilon_i \right] \quad (11)$$

Considering the general form, according to HM, of the internal partition function (Eq. (2)), for a single element, (keeping in mind that $\omega_i$ is unity for the Boltzmann distribution and no particle interaction) and using the relation between the occupation numbers and the internal partition function one can write

$$\frac{Q_{int}(\{N_i\},V,T)}{N} = \frac{g_i \omega_i(\{N_i\},V,T) e^{-\varepsilon_i/K_B T}}{N_i} \quad (12)$$

Equation (12) is used herein to represent both of the cases of high-density gas with the inclusion of $\omega_i(\{N_i\},V,T)$ and the case of low-density gas with excited states (following Boltzmann distribution and no particle interaction) where again for the latter case $\omega_i$ is taken to be unity. Adopting the assumption of uncoupling of the translational energy made by HM [2] and Potekhin [4] and substituting from Eq. (12) into Eq. (11) using the expression for the translational partition function:

$$Q_{trans} = G T^{3/2} V = \left( \frac{2 \pi m K_B}{h^2} \right)^{3/2} T^{3/2} V \quad (13)$$

one gets for the entropy of the gas with excited states the expression

$$S = K_B N \left[ \frac{3}{2} \ln T + \ln G + \ln V + \frac{5}{2} \right] - K_B \sum_i N_i \ln N_i + K_B \sum_i N_i \ln g_i \omega_i(\{N_i\},V,T) \quad (14)$$

The corresponding correct form of the free energy function, is therefore
$$F = U - TS$$
$$= -NK_B T \left[ \frac{3}{2} \ln T + \ln G + \ln V + 1 \right] + \sum_i N_i \varepsilon_i + K_B T \sum_i N_i \ln N_i \quad (15)$$
$$- K_B T \sum_i N_i \ln g_i \omega_i(\{N_i\},V,T)$$

The final forms in Eq. (14) and Eq. (15) can now be easily compared with the entropy and free energy expressions given by HM (Eq. 2.11 & Eq. 2.13 Ref. [2]), for the low-density case (following Boltzmann distribution and no particle interaction), where now it is clear that the last terms in Eq. (14) and Eq. (15) were not included in the HM work (for the duration of this comparison, one keeps in mind that $\omega_i=1$ for all species). Thus, for this latter interpretation, the set of Eqs. (2-11) to (2-13) in the HM work is not self-consistent too.

To summarize, for Eq. (6) to reconcile or to be compatible with Eq. (7), one needs either to remove the excitation energies from Eq. (7) and Eq. (9) where the equations in this case will correctly and consistently represent the idealization of a monatomic perfect gas, or instead to modify Eq. (6) and Eq. (9) as given by Eq. (14) and Eq. (15) (with $\omega_i=1$) to represent an excited monatomic gas following Boltzmann distribution with no particle interaction (note that it can not be called a monatomic perfect gas in this case).

It has also to be noted that the above discussion and the above derived expression for the entropy (see Eq. (11) for example) are in agreement with the discussion in section 46 (Landau and Lifishtiz [29]) and with Eq. (3.48) given by Zel'dovich & Raizer [36] with the definition of an average excitation energy $\bar{\varepsilon}$ such that $N\bar{\varepsilon} = \sum_i N_i \varepsilon_i$. Besides, if one recollects the terms in Eq. (15) again, the basic statistical-thermodynamic relation for the free energy of any state of matter is obtained, that is

$$F = -NK_BT \left(\ln \frac{Q_{tot}}{N} + 1\right) \tag{16}$$

However, it should be remembered that a proper representation of $Q_{int}$ for the material under consideration (e.g. low or high density regimes) has to be taken into account.

At the moment, if one adds to the low-density free energy function, $F_{id}$, (Eq. (15) with $\omega_i=1$, the configurational component $f(V,T,\{N_i\},)$ that depends explicitly on the occupation numbers $\{N_i\}$, the equilibrium condition for an excitation/de-excitation process of the type ($\alpha \leftrightarrow \alpha'$), taking $f$ into consideration, is therefore

$$\frac{\partial(F_{id}+f)}{\partial N_\alpha} = \frac{\partial(F_{id}+f)}{\partial N_{\alpha'}} = C \tag{17}$$

or

$$\varepsilon_\alpha + K_BT\left(1 + \ln(N_\alpha/g_\alpha)\right) + \frac{\partial f}{\partial N_\alpha} = C \tag{18}$$

where $C$ is a constant. Applying the same condition ($\partial F/\partial N_\alpha = \partial F/\partial N_{\alpha'} = C$) to the high density form of the free energy function (Eq. 15) in which the occupational probabilities, $\omega_i$, are used, one gets

$$\varepsilon_\alpha + K_BT\left(1 + \ln(N_\alpha/g_\alpha)\right) + \frac{\partial}{\partial N_\alpha}\left(-K_BT\sum_i N_i \ln \omega_i(\{N_i\},V,T)\right) = C \tag{19}$$

Comparing Eq. (18) and Eq. (19) reveals that

$$f = -K_BT \sum_i N_i \ln \omega_i(\{N_i\},V,T) \tag{20}$$

Upon differentiating (20) with respect to $N_\alpha$ and summing over all excited species we arrive at

$$\sum_\alpha N_\alpha \frac{\partial f}{\partial N_\alpha} = \underbrace{-K_BT \sum_\alpha N_\alpha \ln \omega_\alpha(\{N_i\},V,T)}_{f} - K_BT \sum_\alpha N_\alpha \sum_i N_i \frac{\partial}{\partial N_\alpha} \ln \omega_i(\{N_i\},V,T) \tag{21}$$

or

$$\left[f - \sum_\alpha N_\alpha \frac{\partial f}{\partial N_\alpha}\right] = K_BT \sum_\alpha N_\alpha \sum_i N_i \frac{\partial}{\partial N_\alpha} \ln \omega_i(\{N_i\},V,T) \tag{22}$$

which is the residual term obtained by HM in Ref. [4]. Apparently, the vanishing of this *residual term* would not be achieved unless a restriction is imposed implying that

$$\sum_\alpha N_\alpha \sum_i N_i \frac{\partial}{\partial N_\alpha} \ln \omega_i(\{N_i\},V,T) = 0 \tag{23}$$

or equivalently that

$$\sum_\alpha N_\alpha \frac{\partial f}{\partial N_\alpha} = f \tag{24}$$

Equation (22) gives the residual term $[f - \sum n_i(\partial f / \partial n_i)]$ in terms of its equivalent in the occupational probability domain while Eqs. (23) and (24) express the HM approximate linearization in both domains. It is quite clear from Eq. (24) that $f$ must be linear in $\{N_\alpha\}$ and from Eq. (23) that $\omega_i$ is *independent from* $\{N_\alpha\}$, a restriction that can not easily be satisfied without difficult-to-justify approximations on top of wiping out the basic requirement of the dependence of $\omega_i$ on $\{N_\alpha\}$. Hence, the HM occupation probability formalism does not assure the finiteness of the internal partition function or the manifestation of the phenomenon of pressure ionization at high density on top of being subject to all kinds of criticism against density-independent internal partition functions.

It has also to be noted that even for the simplest way to treat interactions among neutral particles, using the hard sphere model, in which a definite radius is assigned to each particle species, the condition of vanishing the residual term, i.e. $[f - \sum_\alpha N_\alpha \partial f / \partial N_\alpha] = 0$ as required by HM formulation is not satisfied; rather a value $-f$ was obtained. The inclusion of a similar term to account for interactions among charged particles, which depend on the long range Coulomb forces, is another added inaccuracy in the HM formulation. In order to go around these difficulties Hummer & Mihalas had to make every possible approximation to linearize the dependence of interaction terms in the free energy function on $\{N_\alpha\}$. As stated in their paper, the HM linearization leads to an unavoidable double counting with an exponent of the occupation probability smaller by a factor of 2. It has to be recalled, however, that the occupational probability introduced by Hummer & Mihalas does not exclusively assure the convergence or the finiteness of the internal partition function.

In conclusion, any separable part of the configurational component of the free energy is equivalent to a factor (that does not explicitly depend on the occupation numbers of the excited states) to be multiplied by other components of the internal partition function, however, it has to be included once either as a factor in the internal partition function or as a separable part of the configurational component of the free energy. In the latter case, the partition function included in the free energy expression will not be the complete or absolute internal partition function (in which the configurational factor is included) but rather a scaled one. This can be seen easily from Eqs. (15) or (16), however, a detailed explanation of this fact is provided in the appendix . In all cases, the complete or the scaled partition functions need to be finite and require a scheme to truncate the sum over the bound states.

With the HM and similar occupation probability formalisms, one is left with two possibilities; 1- the use of the occupational probability as in Eq. (3) and accurate nonlinear interaction term $F_{config}$ which is statistical-mechanically inconsistent, or 2- the use of the occupational probability as in Eq. (3) and linear interaction term $F_{config}$ which is consistent but inaccurate and does not satisfy the basic requirement of the dependence of $\omega_i$ on the occupation numbers of the excited states, therefore no guarantee for the truncation of the IPF in a neutral gas in such a case.

In the following section we consider the general case of using any scheme to truncate the internal partition function and provide a more accurate and consistent solution of the problem.

## 4– A Consistent Formulation and the Solution of the Inverse Problem

Considering a mixture of chemical elements $j$s at temperatures sufficiently high for all molecules to be fully dissociated, the mixture can therefore be regarded as a mixture of monatomic gases with ions of different ionization multiplicities $\zeta$s in addition to free electrons.

At high densities, degeneracy of the electron gas has to be taken into account; however, ions remain classical because of their heavy masses. Taking the state of the neutral gas to be the standard or reference state, the free energy function for this system can, therefore, be written as

$$F = \underbrace{-\left(2K_B TV / \Lambda_e^3\right) I_{3/2}(\mu_{e,id}/K_B T) + N_e \mu_{e,id}}_{F_{e,id}^{dgc}}$$

$$+ \sum_j \sum_\zeta N_{j,\zeta} \left[ \sum_{m=1}^{\zeta} \chi_{j,m-1} - K_B T \left[ 1 + \ln V - \ln N_{j,\zeta} - \ln \Lambda_{j,\zeta}^3 + \ln \overline{Q}_{j,\zeta}^{int}(\{N_{j,\zeta,i}\},V,T) \right] \right] \quad (25\text{-a})$$

$$+ F_C$$

where $\chi_{j,m}$ is the ionization energy of the ion *j,m* and the term $\sum_{m=1}^{\zeta} \chi_{m-1}$ represents the accumulated sum of the successive ionization energies spent in creating the ground-state of ion $\zeta$ from the ground-state of its neutral atom (reference state); that is equivalent to the energy of ground state of the ion $\zeta$ measured from the ground state level of its precursor neutral atom. In Eq. (25-a), $F_C$ represents a separable part of the configurational component of the free energy that is not exclusively sufficient for the termination of the IPF, if any. One has to note that the partition function $\overline{Q}_{j,\zeta}^{int}(\{N_{j,\zeta,i}\},V,T)$ in Eq. (25-a) is a scaled partition function which in conjunction with the separable part of the configurational free energy, $F_C$, constitutes the complete or absolute internal partition function $Q_{j,\zeta}^{int}(\{N_{j,\zeta,i}\},V,T)$ as explained above. All the results of the discussion given below in regard to the scaled internal partition function also apply to the complete or absolute IPF, where in the latter case $F_C$ is understood to be implicitly embedded in the truncated IPF.

The first term on the right hand side of Eq. (25-a) represents the free energy of an ideal, partially degenerate Fermi electron gas where $\Lambda_k = h / \sqrt{2\pi m_k K_B T}$ is the average thermal wave length of the particle *k* in the mixture, $\mu_{e,id}$ is the chemical potential of the ideal Fermi electron gas and $I_v$ is the complete Fermi-Dirac integral of the order *v*;

$$I_v(x) = \frac{1}{\Gamma(v+1)} \int_0^\infty \frac{y^v}{\exp(y-x)+1} dy \quad (26)$$

where $\Gamma$ is the gamma function. The electron chemical potential $\mu_{e,id}$ is related to the number of free electrons in the system by

$$N_e = \left(2V/\Lambda_e^3\right) I_{1/2}(\mu_{e,id}/K_B T) \quad (27)$$

Without any loss of generality one can write the free energy of a partially degenerate Fermi electron gas as the sum of the classical free energy for electrons, $F_{e,id}^{clc}$, plus a correction term $\Delta F_{e,id}^{dgc}$ that is formally equivalent or equal to the difference between the quantum and classical electrons' free energies; i.e. $\Delta F_{e,id}^{dgc} = \left(F_{e,id}^{dgc} - F_{e,id}^{clc}\right)$. Using the expression for the classical free energy for an ideal electron gas, Eq. (25-a) can now be written as

$$F = \underbrace{- K_B T N_e \left[ 1 + \ln V - \ln N_e - \ln \Lambda_e^3 + \ln 2 \right]}_{F_{e,id}^{cle}}$$

$$+ \sum_j \sum_\zeta N_{j,\zeta} \left[ \sum_{m=1}^{\zeta} \chi_{j,m-1} - K_B T \left[ 1 + \ln V - \ln N_{j,\zeta} - \ln \Lambda_{j,\zeta}^3 + \ln \overline{Q}_{j,\zeta}^{int}(\{N_{j,\zeta,i}\},V,T) \right] \right] \quad (25\text{-b})$$

$$+ \left( \Delta F_e^{dgc} + F_C \right)$$

Various schemes for establishing a finite bound-state partition function are taken into account in Eqs. (25-a) and (25-b) by considering the dependence of the internal partition function on the occupational numbers as well as on volume and temperature. At equilibrium, the free energy is minimum and the occupation numbers can be found by minimizing the free energy in the ionization processes giving rise to a set of ($\sum_j Z_j$) equations of the form

$$\frac{\partial F}{\partial N_{j,\zeta}} - \frac{\partial F}{\partial N_{j,\zeta+1}} - \frac{\partial F}{\partial N_e} = 0 \qquad \zeta = 0,1,\ldots,Z_j \ \& \ j=1,\ldots,J \qquad (28)$$

where $Z_j$ is the atomic number of the element $j$ in the mixture. The minimization condition (28) gives rise to

$$K_B T \ln\left( \frac{N_{j,\zeta}}{N_{j,\zeta+1}} \frac{2 V \overline{Q}_{j,\zeta+1}^{int}(\{N_{j,\zeta,i}\},V,T)}{N_e \Lambda_e^3 \overline{Q}_{j,\zeta}^{int}(\{N_{j,\zeta,i}\},V,T)} \right) + \chi_{j,\zeta} + \underbrace{\left( \frac{\partial}{\partial N_{j,\zeta}} - \frac{\partial}{\partial N_{j,\zeta+1}} - \frac{\partial}{\partial N_e} \right) F_C}_{\Delta \chi_{j,\zeta}^{F_C}}$$

$$+ \underbrace{\left( \frac{\partial}{\partial N_{j,\zeta}} - \frac{\partial}{\partial N_{j,\zeta+1}} - \frac{\partial}{\partial N_e} \right) \Delta F_{e,id}^{dgc}}_{\Delta \chi^{dgc}} - \underbrace{K_B T \sum_{k=1}^{J} \sum_{z=0}^{Z_k} N_{k,z} \left( \frac{\partial}{\partial N_{j,\zeta}} - \frac{\partial}{\partial N_{j,\zeta+1}} - \frac{\partial}{\partial N_e} \right) \ln \overline{Q}_{k,z}^{int}(\{N_{j,\zeta,i}\},V,T)}_{\Delta \chi_{j,\zeta}^{Q^{int\omega}}} = 0$$

$$\zeta = 0,1,\ldots,Z_j \ \& \ j=1,\ldots,J$$

(29)

In Eq. (29), use has been made of the fact that the excitation energies of the two successive ions $j,\zeta$ and $j,\zeta+1$ must be referenced to the same zero level, for which one may choose the ground state of the ion $j,\zeta$ giving rise to the appearance of the ionization energies $\chi_{j,\zeta}$ in the above set of equations. Noting that $\Delta F_{e,id}^{dgc}$ is independent of the occupation numbers $N_{j,\zeta}$s, the following substitutions can be used to ease and improve the readability of Eq. (29) and subsequent Equations:

$$\Delta \chi_{j,\zeta}^{F_C} = \left( \frac{\partial}{\partial N_{j,\zeta}} - \frac{\partial}{\partial N_{j,\zeta+1}} - \frac{\partial}{\partial N_e} \right) F_C$$

$$\Delta \chi^{dgc} = -\frac{\partial \Delta F_{e,id}^{dgc}}{\partial N_e} = K_B T \left( I_{1/2}^{-1}\left(\frac{N_e \Lambda_e^3}{2V}\right) - \ln\left(\frac{N_e \Lambda_e^3}{2V}\right) \right)$$

$$\Delta \chi_{j,\zeta}^{\overline{Q}^{int}} = -K_B T \sum_{k=1}^{J} \sum_{z=0}^{Z_k} N_{k,z} \left( \frac{\partial}{\partial N_{j,\zeta}} - \frac{\partial}{\partial N_{j,\zeta+1}} - \frac{\partial}{\partial N_e} \right) \ln \overline{Q}_{k,z}^{int}(\{N_{j,\zeta,i}\},V,T)$$

$$= -K_B T \sum_{k=1}^{J} \sum_{z=0}^{Z_k} \frac{N_{k,z}}{\overline{Q}_{k,z}^{int}(\{N_{j,\zeta,i}\},V,T)} \left( \frac{\partial}{\partial N_{j,\zeta}} - \frac{\partial}{\partial N_{j,\zeta+1}} - \frac{\partial}{\partial N_e} \right) \overline{Q}_{k,z}^{int}(\{N_{j,\zeta,i}\},V,T)$$

$$\zeta = 0,1,....,Z_j \quad \& \quad j = 1,....,J$$

(30)

The last term in the left-hand-side of Eq. (29) (also the last substitution in (30)) is a result of the dependence of the scaled internal partition functions on the occupation numbers of the species; that is a consequence of the scheme used to effectively terminate the internal partition function. This term which is frequently ignored by many authors in the literature is necessary for the thermodynamic consistency of the formulation. We have seen above that forcing this term to vanish, when using occupational probabilities similar to the HM formulation, implied a density-independent scaled IPF demolishing a basic constraint set for the IPF to depend on the occupation numbers in order to assure finiteness of the IPF. This result can be generalized for any scheme used to truncate the IPF as shown below, that is to say, a separable configurational component of the free energy (or the vanishing of this residual term) implies a density-independent IPF with all of the above-mentioned associated shortcomings. Since the occupation numbers and the IPFs are always nonnegative, therefore the vanishing of the residual term as it appears in Eq. (30) requires

$$\frac{\partial}{\partial N_{j,\zeta}} \overline{Q}_{k,z}^{int}(\{N_{j,\zeta,i}\},V,T) = \left( \frac{\partial}{\partial N_{j,\zeta-1}} - \frac{\partial}{\partial N_e} \right) \overline{Q}_{k,z}^{int}(\{N_{j,\zeta,i}\},V,T) \tag{31}$$

$$\zeta = 1,2,....,Z_j \quad \& \quad j = 1,....,J$$

Successive use of the relation (31) gives

$$\frac{\partial \overline{Q}_{k,z}^{int}}{\partial N_{j,\zeta}} = \frac{\partial \overline{Q}_{k,z}^{int}}{\partial N_{j,0}} - \zeta \frac{\partial \overline{Q}_{k,z}^{int}}{\partial N_e} \qquad \zeta=1,2,..., Z_j \ \& \ j=1,....,J \tag{32}$$

Applying the constraint of electro-neutrality, the differential of the IPF can be expressed as

$$d\overline{Q}_{k,z}^{int} = \sum_{j} \sum_{\zeta=0}^{Z_j} \left( \frac{\partial \overline{Q}_{k,z}^{int}}{\partial N_{j,\zeta}} + \zeta \frac{\partial \overline{Q}_{k,z}^{int}}{\partial N_e} \right) dN_{j,\zeta} + \frac{\partial \overline{Q}_{k,z}^{int}}{\partial T} dT + \frac{\partial \overline{Q}_{k,z}^{int}}{\partial V} dV \tag{33}$$

Assuming that all chemical reactions other than ionization and recombination are neglected, which is the case for a mixture of inert gases or a plasma mixture at temperatures sufficiently high for polyatomic molecules to be considered as fully dissociated. In such a case the number of heavy particles of any chemical element, $N_{h,j}$, and the total number of heavy particles in the system $N_H = \sum_j N_{h,j}$ remain unchanged (i.e; $dN_{h,j}=0$ and $dN_H=0$). Upon substitution from Eq. (32) into Eq. (33) and using the last stoichiometric constraint of conservation of elemental nuclei, one gets

$$\begin{aligned}
d\overline{Q}_{k,z}^{int} &= \sum_j \frac{\partial \overline{Q}_{k,z}^{int}}{\partial N_{j,0}} dN_{j,0} + \sum_j \sum_{\zeta=1}^{Z_j} \left(\frac{\partial \overline{Q}_{k,z}^{int}}{\partial N_{j,0}}\right) dN_{j,\zeta} + \frac{\partial \overline{Q}_{k,z}^{int}}{\partial T} dT + \frac{\partial \overline{Q}_{k,z}^{int}}{\partial V} dV \\
&= \sum_j \frac{\partial \overline{Q}_{k,z}^{int}}{\partial N_{j,0}} dN_{j,0} + \sum_j \frac{\partial \overline{Q}_{k,z}^{int}}{\partial N_{j,0}} \sum_{\zeta=1}^{Z_j} dN_{j,\zeta} + \frac{\partial \overline{Q}_{k,z}^{int}}{\partial T} dT + \frac{\partial \overline{Q}_{k,z}^{int}}{\partial V} dV \\
&= \sum_j \frac{\partial \overline{Q}_{k,z}^{int}}{\partial N_{j,0}} \underbrace{\sum_{\zeta=0}^{Z_j} dN_{j,\zeta}}_{dN_{h,j}=0} + \frac{\partial \overline{Q}_{k,z}^{int}}{\partial T} dT + \frac{\partial \overline{Q}_{k,z}^{int}}{\partial V} dV \\
&= \frac{\partial \overline{Q}_{k,z}^{int}}{\partial T} dT + \frac{\partial \overline{Q}_{k,z}^{int}}{\partial V} dV
\end{aligned} \quad (34)$$

The final result of Eq. (34) indicates that for the case of a separable configurational component of the free energy (vanishing of the residual term), the thermodynamically consistent IPF will be independent of the occupation numbers with no guarantee of the finiteness of the IPF.

Apart from the issue of the separability of the configurational component of the free energy, the solution of the set of nonlinear equations (29) will give the occupation numbers of all particle species in the mixture given that a scheme for establishing finite bound-state partition functions is known or adopted in advance. Using the above substitutions in (30), the condition (29) therefore leads to a set of coupled nonlinear equations of the form

$$\frac{N_{j,\zeta+1} N_e}{N_{j,\zeta}} = 2V\Lambda_e^3 \frac{\overline{Q}_{j,\zeta+1}^{int}(\{N_{j,\zeta,i}\},V,T)}{\overline{Q}_{j,\zeta}^{int}(\{N_{j,\zeta,i}\},V,T)} \times exp\left(\frac{-\left[\chi_{j,\zeta} - \Delta\chi_{j,\zeta}^{F_c} - \Delta\chi^{dgc} - \Delta\chi_{j,\zeta}^{\overline{Q}^{int}}\right]}{K_B T}\right) \quad \zeta=0,1,....,Z_j \ \& \ j=1,....,J \quad (35)$$

The scaled internal partition functions in Eqs. (35) are those forms adopted in advance ensuring the convergence or finiteness of the IPFs. The solution of the set of Eqs (35) subjected to the constraints of electroneutrality and conservation of nuclei gives the equilibrium occupation numbers of all species. Noting that the set of Eqs. (35) have the form of corrected or modified Saha equations, available algorithms for the solution of Saha equations can be used to calculate the population numbers with relative simplicity and high accuracy. It may be enlightening to note here that the residual term $\Delta\chi_{j,\zeta}^{\overline{Q}^{int}}$ appears in Eq. (35) as a correction or modification of the ionization energy. Accordingly, the integrability criterion derived in a previous work [28] can be simply applied to this term to show whether the scheme used to truncate the IPF can be directly related to a separable configurational component of the free energy function, or not. It will be seen therefore that *a scheme capable to truncate the IPF can never be fully condensed exclusively in a separable analytic and continuous expression for the configurational component*.

The remaining part to complete the consistent formulation of the problem is therefore to derive the modifications to the set of thermodynamic functions corresponding to the adopted scheme to establish a converging or finite IPF. This task can easily be performed by differentiating Eq. (25-a) directly where the pressure is given by

$$P = -\left(\frac{\partial F}{\partial V}\right)_{T,\{N_{j,\zeta,i}\}} = (2K_B TV/\Lambda_e^3) I_{3/2}(\mu_{e,id}/K_B T) + \frac{N_H K_B T}{V} - \left(\frac{\partial F_C}{\partial V}\right)_{T,\{N_{j,\zeta,i}\}}$$
$$+ K_B T N_H \sum_j \sum_\zeta \alpha_{j,\zeta} \frac{\partial}{\partial V} \ln \overline{Q}_{j,\zeta}^{int}(\{N_{j,\zeta,i}\},V,T) \tag{36}$$

Similarly, the internal energy can be expressed as

$$U = -T^2 \left(\frac{\partial F/T}{\partial T}\right)_{V,\{N_{j,\zeta,i}\}}$$
$$= (3K_B TV/\Lambda_e^3) I_{3/2}(\mu_{e,id}/K_B T) + \frac{3N_H K_B T}{2V} - T^2 \left(\frac{\partial (F_C/T)}{\partial T}\right)_{V,\{N_{j,\zeta,i}\}} \tag{37}$$
$$+ N_H \sum_j \sum_\zeta \sum_{m=1}^{\zeta} \alpha_{j,\zeta} \chi_{j,m-1} + K_B T^2 N_H \sum_j \sum_\zeta \alpha_{j,\zeta} \frac{\partial}{\partial T} \ln \overline{Q}_{j,\zeta}^{int}(\{N_{j,\zeta,i}\},V,T)$$

The excitation energies of all species are embedded in the last term in Eq. (37). The set of Eqs (35, 36 and 37) in conjunction with the constraints of electroneutrality and conservation of nuclei together with the scheme used to establish finite bound-state partition functions represent the complete set of equations to be used to consistently calculate the occupation numbers and thermodynamic properties of the nonideal plasma system under consideration, within the chemical picture.

## V. ILLUSTRATIVE EXAMPLE

In section **4** we presented a consistent mathematical formulation of the problem of the calculation of thermodynamic properties and the establishment of finite bound state partition functions for a multi-component plasma mixture within the chemical picture. The formulation is a generalization of a previously introduced formulation for the case of hydrogen plasma [32]. As has been shown above, the consistent establishment of finite IPFs in the inverse scheme necessitates the introduction of new correction terms to the ionization potentials, $\Delta \chi_{j,\zeta}^{\overline{Q}^{int}}$, plasma pressure, $\Delta P_{Q_w} = K_B T N_H \sum_j \sum_\zeta \alpha_{j,\zeta} \frac{\partial}{\partial V} \ln \overline{Q}_{j,\zeta}^{int}(\{N_{j,\zeta,i}\},V,T)$, and internal energy, $\Delta U_{Q_w} = K_B T^2 N_H \sum_j \sum_\zeta \alpha_{j,\zeta} \frac{\partial}{\partial T} \ln \overline{Q}_{j,\zeta}^{int}(\{N_{j,\zeta,i}\},V,T)$, as consequences of the scheme used, in advance, to truncate the IPF. In this section, an illustrative example is worked out to show both simplicity and effectiveness of the proposed scheme and the significance of including these corrections on top of being essential for consistency.

The present example is worked out for a 15-element solar mixture using the same models for the occupational probability and Coulombic interaction term used in Ref. [32]. As indicated in Ref [32], in a more rigorous theory, the occupational probability and the interaction terms may be derived based on the same physical model of particle interactions. It may be recalled that the only consequence of using the occupational probability as in Ref [32] is the correction to the pressure $\Delta P_{Q_w}$ or the pressure of shell compression.

Figure 1 shows the individual components of the pressure due to free electrons $P_e$, Coulombic correction $\Delta P_{F_C}$, shell compression $\Delta P_{Q_\omega}$ and the total pressure of a 20000 K, 15-element solar mixture plasma all normalized to $nK_BT$ where $n$ is the number density of heavy particles. The component of the pressure due heavy particles (not shown in the figure) is understood to be unity in this case. As expected, the Coulombic correction term (separable part of the configurational free energy) contributes a negative component to the pressure which becomes significant at high densities. The component of the free electrons' pressure increases significantly at high densities, as the degree of degeneracy increases, preventing the collapse of the plasma in this case as explained in Ref. [32]. As it can be seen from the figure, the shell compression component $\Delta P_{Q_\omega}$ has a non-negligible effect and contributing a significant positive component of the computed total pressure. This component is however essential for the self-consistency of the set of thermodynamic functions along with the calculated occupation numbers. Figure 2 shows the total pressure for the same plasma mixture with and without the shell compression component $\Delta P_{Q_\omega}$ confirming the quantitative importance of this component as discussed above.

For completeness we provide in Figures 3 and 4, respectively a set of isotherms of the pressure and degree of ionization of the 15-element solar mixture as functions of the heavy particle number density.

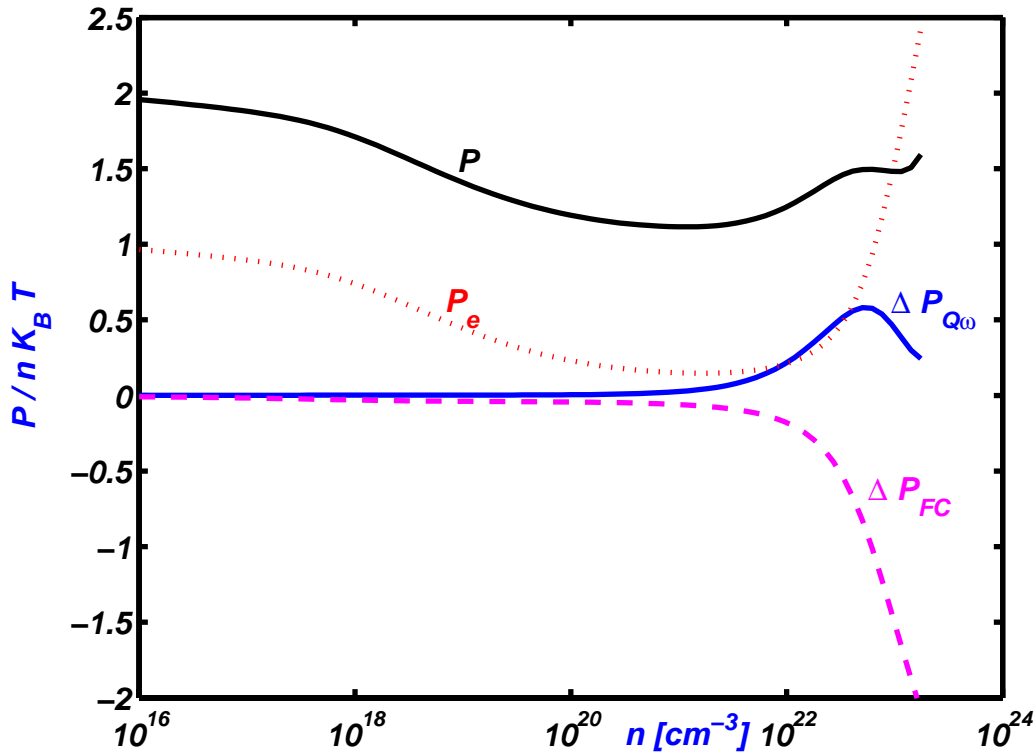

**Figure 1** Normalized components of free electrons' pressure, $P_e$, Coulombic correction term, $\Delta P_{F_C}$, and shell compression term, $\Delta P_{Q_\omega}$, along with the total pressure of a 20000 K, 15-element solar mixture as functions of the number density of heavy particles.

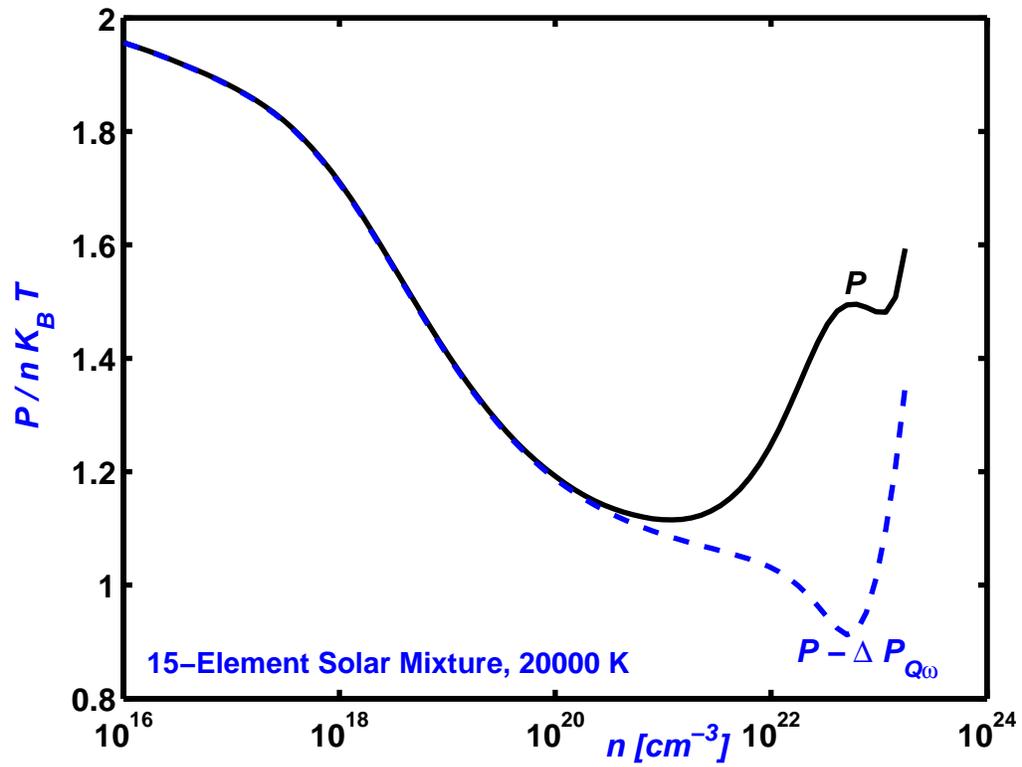

**Figure 2** Normalized total pressure of a 20000 K 15-element solar mixture plasma with and without the shell compression component, $\Delta P_{Q_\omega}$.

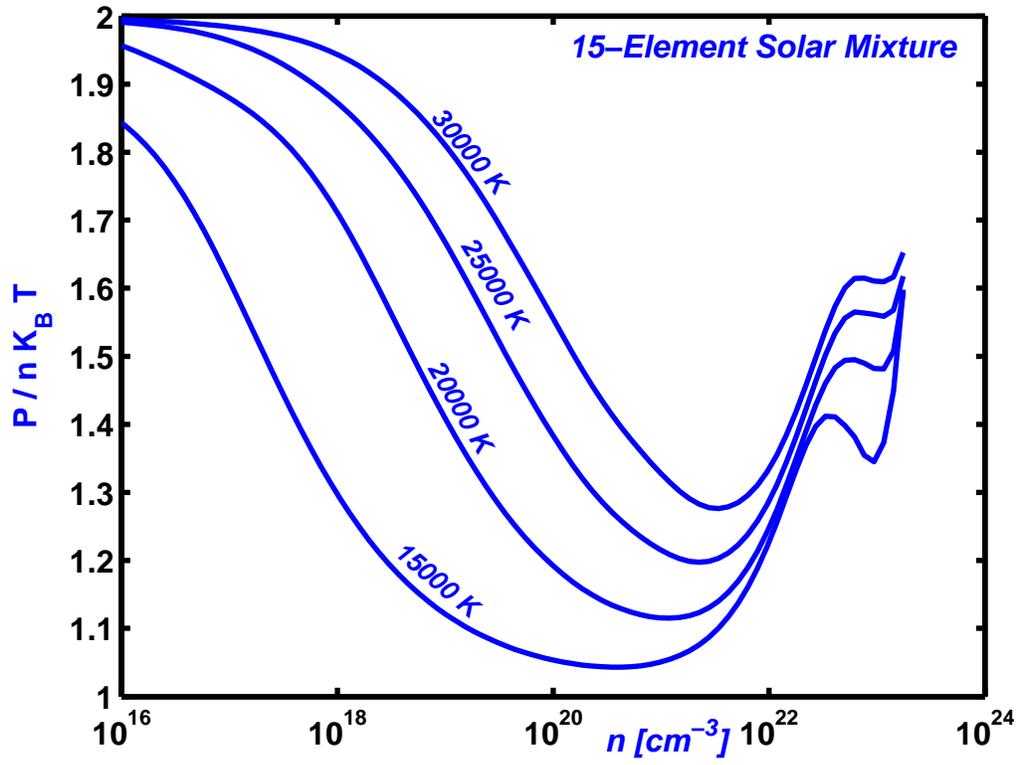

**Figure 3** Isotherms of normalized total pressure of 15-element solar mixture plasma as functions of the heavy particle number density.

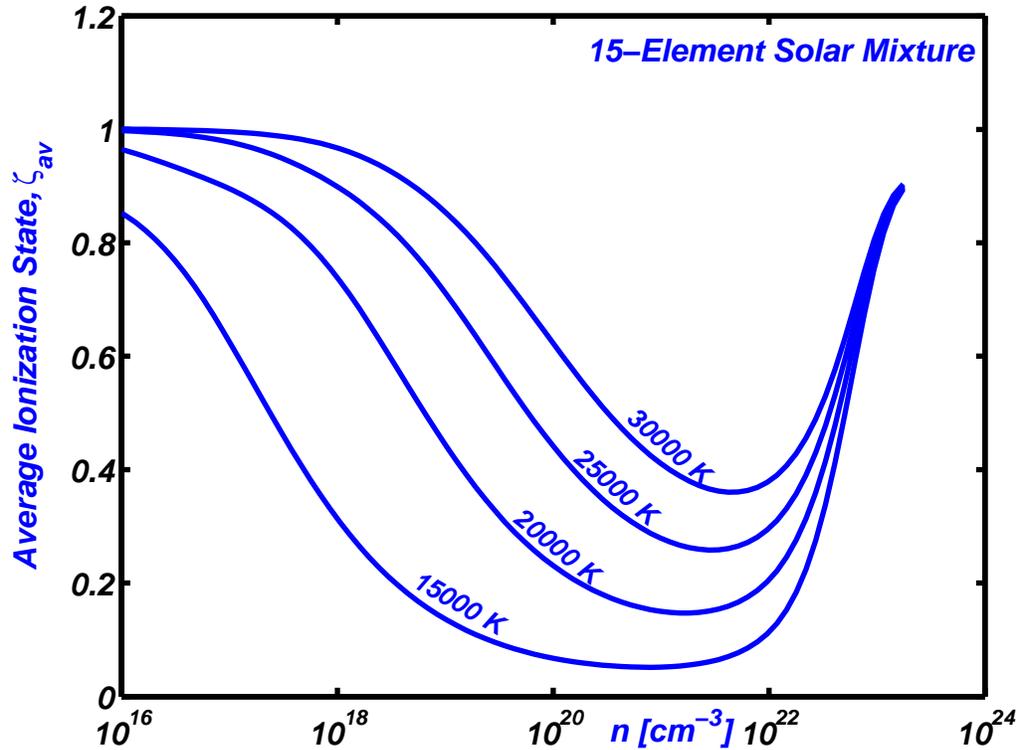

**Figure 4** Isotherms of the degree of ionization of 15-element solar mixture plasma as functions of the heavy particle number density

## 6-Summary and Conclusions

The problem of thermodynamically-consistent calculation of equilibrium thermodynamic properties and the establishment of finite bound-state partition functions in nonideal plasma mixtures is scrutinized within the chemical picture. The present exploration clears ambiguities and gives a better understanding of the problem on top of pointing out weaknesses and inaccuracies buried in widely used models in the literature. A consistent formulation of the problem, within the chemical picture, in terms of the solution of the inverse problem is introduced. The proposed scheme generalizes the scheme previously introduced for hydrogen plasma to the case of nonideal multi-component plasma mixtures. A nontrivial example is worked out showing the quantitative significance of including those correction terms necessary for the thermodynamic consistency terms which have been neglected in the majority of publications in the literature.

### Acknowledgments

The author wishes to thank Dr. W. Metwally from GE (GNF), Wilmington, NC for suggesting improvements to the manuscript and Prof. T. Thiemann from the Chemistry Dept., UAE University for the time and efforts he courteously devoted in translating Fermi's original article (reference 15) from German to English.

This work is supported by the UAE University, contract 05-02-2-11/08.

# Appendix

Let $F_C$ in Eq. (25-b) be written as

$$F_C = \frac{-K_B T F_C}{-N_H K_B T} \sum_j \sum_\zeta N_{j,\zeta}$$
$$= -\sum_j \sum_\zeta N_{j,\zeta} K_B T \ln\{\exp(-F_C / N_H K_B T)\} \quad (A-1)$$

This is equivalent to multiplying $\overline{Q}_{j,\zeta}^{int}$ in Eq. (25-b) by the factor $\exp(-F_C / N_H K_B T)$. Now, in Eq. (29) the third term in the left hand side, namely,

$$(\partial/\partial N_{j,\zeta} - \partial/\partial N_{j,\zeta+1} - \partial/\partial N_e) F_C, \quad \zeta = 0,1,....Z_j \,\&\, j = 1,....,J \quad (A-2)$$

should be recovered or obtained from the last term in the left hand side if we multiply $\overline{Q}_{j,\zeta}^{int}$ by the factor $\exp(-F_C / N_H K_B T)$. Note that multiplying $\overline{Q}_{j,\zeta}^{int}$ by this factor in the first term in Eq. (29) produces no change in the term since $\overline{Q}_{j,\zeta}^{int}$ appears both in the numerator and denominator of the argument of the logarithmic term. For the last term in the left hand side of Eq. (29), multiplying $\overline{Q}_{j,\zeta}^{int}$ by the above mentioned factor is equivalent to adding a new term (noting that $ln(ab) = ln(a) + ln(b)$) given by

$$-K_B T \sum_{k=1}^{J} \sum_{z=0}^{Z_k} N_{k,z} (\partial/\partial N_{j,\zeta} - \partial/\partial N_{j,\zeta+1} - \partial/\partial N_e) \ln\{\exp(-F_C / N_H K_B T)\},$$
$$\zeta = 0,1,....Z_j \,\&\, j = 1,....,J \quad (A-3)$$

which can be written as

$$\sum_{k=1}^{J} \sum_{z=0}^{Z_k} N_{k,z} (\partial/\partial N_{j,\zeta} - \partial/\partial N_{j,\zeta+1} - \partial/\partial N_e)(F_C / N_H),$$
$$\zeta = 0,1,....Z_j \,\&\, j = 1,....,J \quad (A-4)$$

Since $F_C$ does not depend on $k$ or $z$ (summation indices) while the partial differentiation is to be performed with respect to $N_{j,\zeta}$, $N_{j,\zeta+1}$ and $N_e$, therefore one can take $(\partial/\partial N_{j,\zeta} - \partial/\partial N_{j,\zeta+1} - \partial/\partial N_e)(F_C / N_H)$ outside the summation as a common factor with the result that the above term becomes

$$(\partial/\partial N_{j,\zeta} - \partial/\partial N_{j,\zeta+1} - \partial/\partial N_e)(F_C / N_H) \underbrace{\sum_{k=1}^{J} \sum_{z=0}^{Z_k} N_{k,z}}_{N_H} = (\partial/\partial N_{j,\zeta} - \partial/\partial N_{j,\zeta+1} - \partial/\partial N_e) F_C,$$
$$(A-5)$$

$\zeta = 0,1,....Z_j \,\&\, j = 1,....,J$

which is the third term in the left hand side of Eq. (29). Accordingly, any separable part of the configurational component, $F_C$, of the free energy (factorizable configurational component of the partition function) is equivalent to a factor given by $\omega = \exp(-F_C / N_H K_B T)$ to be multiplied by other components of the internal partition function, however, it has to be included once either as a factor in the IPF or as a separable part of the configuration component of the free energy. Apparently this factor cannot exclusively truncate the IPF as it can be factored out from the sum over the states and the need to truncate the sum persists.